# Minimum quench power dissipation and current non-uniformity in ITER type NbTi cable-in-conduit conductor samples under DC conditions

**G Rolando, E P A van Lanen, A Nijhuis**

Energy, Materials and Systems, Faculty of Science and Technology, University of Twente, P.O. Box 217, 7500 AE Enschede, Netherlands

E-mail: g.rolando@utwente.nl, e.p.a.vanlanen@utwente.nl, a.nijhuis@utwente.nl

**Abstract**. The level of current non-uniformity in NbTi CICCs sections near the joints in combination with the magnet field profile needs attention in view of proper joint design. The strand Joule power and current distribution at quench under DC conditions of two samples of ITER Poloidal Field Coil conductors, as tested in the SULTAN facility and of the so called PFCI Model Coil Insert, have been analyzed with the numerical cable model JackPot. The precise trajectories of all individual strands, joint design, cabling configuration, spatial distribution of the magnetic field, sample geometry and using experimentally determined interstrand resistance distributions have been taken into account. Although unable to predict the quench point due to the lack of a thermal-hydraulic routine, the model allows to assess the instantaneous strand power at quench and its local distribution in the cable, showing the hot spots, once the quench conditions in terms of current and temperature are experimentally known., The analysis points out the relation of the above mentioned factors with the DC quench stability of both short samples and coils. The possible small scale and local electrical-thermal interactions were ignored in order to examine the relevance of such effects in the overall prediction of the CICC performance The electromagnetic code shows an excellent quantitative predictive potential for CICC transport properties, excluding any freedom for matching the results. The influence of the local thermal effects in the modeling is identified as being marginal and far less than the generally accepted temperature margin for safe operation.

1. **Introduction**
Cable-In-Conduit Conductors (CICCs) are commonly used in fusion magnets due to their elevated stability and current carrying potential. The CICCs foreseen for the poloidal field (PF) coils of the fusion reactor ITER will contain hundreds of NbTi strands twisted and compacted together in multiple cabling stages. High currents require large cable diameters and produce a significant self field that, summed to the background field, determines the magnetic field distribution over the CICC cross section. Following their trajectories from one side of the conductor to the opposite, the twisted strands experience large variations of the local magnetic field inside the cable and may reach elevated electric field levels on extremely localised sections. As a result, the power dissipation over the cross section is not homogeneous. On top of that, the currents in the strands of CICCs may not be uniformly distributed. Typical sources of unbalance in strand currents can be found in the joints at the extremities of a conductor. The spread in the electrical resistances between superconducting strands and the normal conducting parts in the terminals of the current leads to the power supply causes a corresponding spread in the currents of the strands. The combination of magnet self field profile and current non-uniformity among the strands results in local hot spots that can lead, in the worst case, to premature quenches. The thermal equilibrium in the CICC, determined by the heat transfer and the He mass flow, significantly affects the stability against quasi- stable perturbations and thus the maximum obtainable current and voltage, after which the take-off evolves irreversible into a quench.

The aim of this work is to explore quantitatively the quench limitations of ITER type NbTi CICC in relation to the joint layout and magnet field map and evaluate the scope for optimisation of the design parameters for magnets. In order to minimize the heat load in joints due to AC losses, high resistivity components are required. This, however, increases the heat load from DC resistance and may raise the current non-uniformity level as well. Current unbalance combined with the large CICC self field gradient might result in locally concentrated dissipation hot spots. Once the allowable local peak power is known, the acceptable level of current non-uniformity in relation to the magnetic field and temperature profile on the conductor can be determined. The JackPot-AC Joint Model, presently under development, can then be used to optimise the electric resistances of the joint components for minimal heat load from AC losses and DC resistance.

To assess the performance of the full size PF conductor, both short length samples and a superconducting insert coil have been manufactured. The Poloidal Field Insert Sample (PFIS) was tested in the SULTAN facility at CRPP (Switzerland) in 2004. The DC behaviour of the sample revealed low performance compared to design expectations due to early quench of the conductor [1]. The cause of this result was attributed to current unbalance introduced by the joints that could not be redistributed before the high field region of the sample, thus causing the overloaded strands to reach saturation and eventually leading to a quench of the entire conductor [2, 3]. An important role in the phenomenon was recognized to be played by the cable self field, producing a considerable magnetic field gradient over the conductor cross section [4, 5]. The uneven power dissipation produced by the simultaneous actions of the above factors can lead in CICC to high electric field values on short strand sections. At the same time, the average electric field along the conductor may remain too low to be experimentally detectable before the onset of the quench. This behaviour is obviously connected to the size of the composite conductor and the strand properties in terms of both variation of the steepness of the Ic versus B curve and broadness of the voltage-current transition (n-value).

In order to get a better understanding of the behaviour of high current NbTi CICCs and joints in relevant ITER operational conditions, the Poloidal Field Conductor Insert (PFCI) was built in Europe and tested at JAERI Naka (Japan) in 2008. Although it was constructed using the same cable as the PFIS, the PFCI DC behaviour met the ITER operation requirements without the premature quenches observed in the SULTAN sample, due to the longer conductor length and increased distance between the joints and the high field region [6]. More recently, the first Chinese PF conductor sample (CNPF1) was tested successfully at the SULTAN facility. The sample featured a new layout for the upper terminations together with a 'U' bend bottom to avoid the current unbalance produced by conventional joints.

In this study, the strand behaviour at quench under DC conditions of the previously mentioned conductors has been analysed with the JackPot code [7]. Being a DC electromagnetic model, without a thermal-hydraulic description of the conductor, JackPot cannot be used to predict the quench point of a CICC or describe the quench characteristics (i.e. occurrence of sudden transitions) and evolution, which is beyond the intention of the present work. However, given the temperature and current conditions characterizing the quench, which are known from the experiment, the code does accurately calculate the voltages and currents in a cable, taking into account the precise trajectories of all individual strands, joint design, cabling configuration, spatial distribution of the magnetic field, sample geometry and experimentally determined interstrand resistance distributions. Therefore the code allows for the determination of the power dissipation, either locally or globally, at the instant at which the measured sample quenched during the experiment. These values set a conservative limit to the allowable heat generation at quench, since JackPot assumes a uniform temperature distribution in the cross-section, while in reality quenches appear in the form of extremely localized hot spots with likely higher temperature than the average cable measured one. However, at this point we assume by approximation that the local thermal-electrical interactions have a minor effect on the overall analysis. To which extent this assumption is justified will follow implicitly from the analysis results. Although it is obvious that the local temperature at the hot-spot is higher than the average cable temperature, the difference in temperature between the measured average and the hot spot cannot be quantified accurately and any existing hypothesis is merely based on the outcome of cable models including thermo-hydraulic descriptions that up to now have not shown the ability for a reasonable quantitative prediction of the take-off point for NbTi CICCs. At this purpose it should be remarked that in spite of the effort spent in the last years in the attempt to model the CICC DC transport behavior, none of the developed models has been able up to now to reach sufficient accuracy in the reconstruction of the V-T curve. While all other existing models are characterized by an accuracy typically of about 0.5 K (depending on the conductor current level), the $T_{cs}$ of all the analyzed samples could be assessed with JackPot within 0.12 K along the entire B and T experimental spectrum [8-10]. We believe that the accuracy and credibility of JackPot can be attributed to taking into account all the individual strands inside a CICC (over lengths of tens of meters) in combination with a very detailed model of the joints based on extensive joint measurements and not including any free parameters that could be tuned to match the experimental results.

The aim of the present work is obviously not to predict the evolution of a quench or the thermal runaway, but to estimate the allowed minimum power dissipation and its dependency from current non-uniformity in PF type of conductors, as a basis for the PF joint design optimisation. For this purpose it is required to exclude as much as possible uncertainties and stay close to the actual runaway conditions determined in relevant experiments. In this work we show that it is possible to reconstruct the V-T curves with high accuracy and as such to find the best possible assessment of the peak power at quench, its location and volume, based on the cable measured temperature profile without allowing any speculative assumptions on local temperature distributions. We believe that this approach is permitted from a pragmatic point of view, taking into account the poor accuracy of the available thermo-hydraulic concepts and the detailed electromagnetic approach with minimum uncertainty in combination with measured cable temperature as presented here.

The analysis shows the effect of the above cited conductor parameters on the DC quench stability of both short samples and coils. In addition to those factors, helium temperature, pressure and flow rate, cable design and void fraction as well as strand properties also play a role on CICC DC performance. However, this paper focuses exclusively on the analysis of the conductors in ITER operating conditions or in the experimental set up employed in the considered tests. Table 1 summarizes some design and operating parameters for the different PF coils [11].

**Table 1.** Design and operating parameters of ITER PF coils.

|  | PF2, 3, 4 | PF5 | PF1,6 |
|---|---|---|---|
| Type of strand | NbTi | NbTi | NbTi |
| Number of SC strands | 720 | 1152 | 1440 |
| Number of Cu strands | 360 | 6 | - |
| Nominal Operating Current [kA] | 50.0 | 52.0 | 48.0 |
| Nominal Effective Peak Field [T] (external + self field) | 5.0 | 5.7 | 6.4 |
| Operating temperature [K] | 4.5 | 4.5 | 4.5 |

2. **Conductor and sample layout**
The preparation and instrumentation details of the PFIS can be found in [12]. The sample comprises two conductor sections ('legs') connected at one end by a hairpin joint. One section (left leg) consists of the regular CICC of the poloidal field coil insert (W) while the second (right leg) had the steel wraps around the final cabling stage sub-cables completely removed before jacketing (NW). The PFIS size is the standard one for SULTAN samples, about 3.5 m long, including the bottom joint and the upper connections. The terminations of the sample were prepared by swaging the cable into CuCrZr sleeves. The bonding between cable and sleeve was obtained by removing the Ni coating from the strands at the cable surface, and replacing it with Ag coating, in combination with tinning of the inner surface of the sleeve. At the bottom joint the sleeves were joined by five copper saddles, while copper plates connected the upper terminations to the secondary winding of the SULTAN transformer. The contact between sleeve and Cu saddles/plates was obtained through indium wires pressed by the joint clamp to a 0.2 mm thick layer. The main parameters for the two legs and the joints are summarized in Table 2 [12, 13].

**Table 2.** PFIS conductor & sample parameters.

|  | PFIS W | PFIS NW |
|---|---|---|
| Number of SC strands | 1440 | |
| Strand diameter [mm] | 0.73 | |
| Cabling sequence | 3 x 4 x 4 x 5 x 6 | |
| Last stage twist pitch [mm] | 489 | 530 |
| Cable outer diameter [mm] | 38 | |
|  | Bottom Joint | Upper Termination |
| Termination length [mm] | 450 | 420 |
| Cu plate RRR | 41 | 86 |
| CuCrZr sleeve RRR | 2.5 | |
| Sleeve – plate contact angle | 135° | |
| Sleeve outer diameter [mm] | 42 | |

The design and manufacture of the PFCI are described in [14-16]. The PFCI is a single-layer solenoid wound out of 50 m of the same NbTi CICC used for the PFIS W (see Figure 1), apart from a slight variation in the twist pitch sequence, see Table 3. The upper section of the main winding was connected to an intermediate joint, to test the joint behaviour in ITER-relevant magnetic field conditions. The PFCI features the same joint/terminations design of the PFIS; however improvements have been made to the RRR of the materials and the strand-sleeve soldering technique as detailed in Table 3.

**Table 3.** PFCI sample modifications compared to PFIS.

|  | PFCI | PFIS W |
|---|---|---|
| Twist pitch sequence [mm] | 45 x 85 x 125 x 160 x 410 | 42 x 86 x 122 x 158 x 489 |
| Cu plate RRR | 300 | 41/86 |
| CuCrZr sleeve RRR | 6 | 2.5 |
| Strand-sleeve soldering | SnPb paste coating | Ag coating + sleeve tinning |

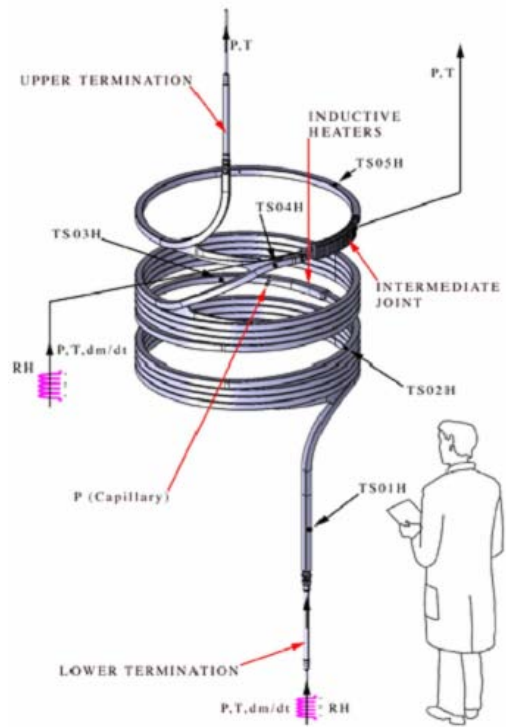

**Figure 1.** View of the PF Conductor Insert [13 - 15].

Unlikely the PFIS and PFCI samples, the PFCN1 constitutes an example of PF2 conductor. Details about its assembly are reported in [17]; while the main parameters of the conductor can be found in Table 4. The most important innovations of the PFCN1 are represented by the new layouts of both upper terminations and bottom joint. For the connections to the SULTAN transformer, a combo box solution developed at CEA was adopted. As shown in Figure 2, wraps are removed at the extremities of the cable, which allows petals to be individually soldered into grooves machined in a copper/stainless steel plate. The bottom joint was replaced by a "U" bend inserted into a hairpin bending box, see Figure 3.

**Table 4.** PFCN1 conductor parameters.

|  | PFCN1 |
|---|---|
| Number of SC strands | 720 |
| SC Strand diameter [mm] | 0.76 |
| Cabling sequence | ((((2SC + 1Cu strand) x 3 x 4 +1 Cu core 1) x 5) + 1 Cu core 2) x 6 |
| Twist pitches [mm] | 45x85x145x250x450 |
| Cable outer diameter [mm] | 35.3 |

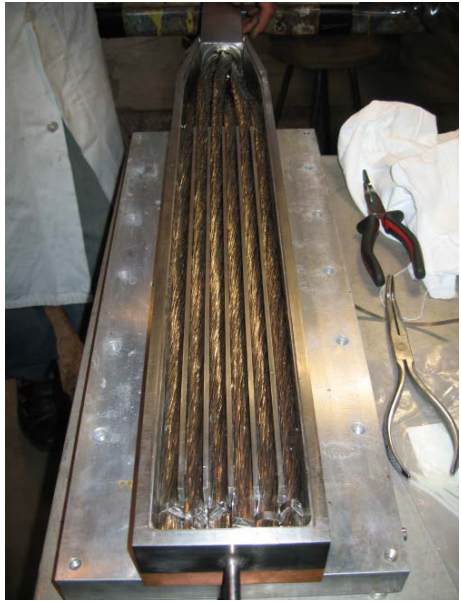

**Figure 2.** PFCN1 upper termination featuring the combo box design, courtesy of L. Reccia [17].

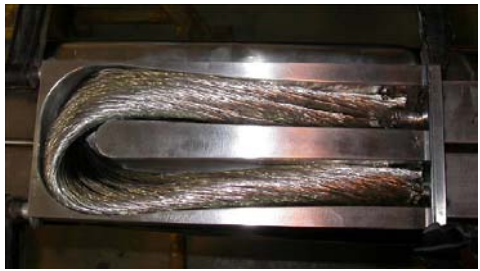

**Figure 3.** PFCN1 bottom"U" bend in the hairpin box, courtesy of L. Bo [16].

### 3. Modelling with JackPot

JackPot is an electrical model for the simulation of the DC behaviour of CICCs [7, 19]. One of the main features of the code consists in making explicit use of the precisely calculated trajectories of all the strands to evaluate key parameters such as the magnetic field profile on the strands, the interstrand contacts and the strand to joint contacts. The latter two distributions are then multiplied by resistivity parameters to obtain the interstrand resistances (parameter ρss) and strand to joint resistances (parameter ρsj). The parameters ρss and ρsj are fixed by matching the results from simulations with interstrand [20] and joint [21] resistance measurements. The possible addition of solder in the cable joints to improve the electrical contact between strands and sleeve can also be taken into account. The soldering procedure results in a double positive effect on current distribution. On the one hand, the number of strands in contact with the sleeve is increased, and on the other hand the strand-sleeve contact area is enlarged. The enhanced contact is equivalently reproduced in JackPot by increasing the diameter of the strands located within the solder layer depth. The complete DC current model for joints has been validated through measurements [22]. Additionally, the current uniformity in CICC is also affected by the presence of stainless steel wraps around the last cabling stage bundles. Wrapping has been proved to increase the resistances between strands from different petals, making it more difficult for current to re-distribute [20]. Therefore, in JackPot the consequent highly impeded current transfer is modelled as absence of contacts between strands located in adjacent petals. The assumption can be regarded as acceptable even in the case of a long coil as the PFCI considering that the highest current unbalance, and thus re-distribution, is expected within petals and not among petals [23]. Moreover, inter-petal current re-distribution is likely to play an increasing role as the course of the quench progresses, while the present analysis focuses on the very beginning of the process. Table 5 summarises the parameter set derived for the PFIS simulation.

**Table 5.** Summary of the JackPot model parameters used in the PFIS simulation.

|  | PFIS W | PFIS NW |
|---|---|---|
| Interstrand resistivity parameter - cable region [x$10^{-12}$ $\Omega \cdot m^2$] | 159 | 101 |
| Interstrand resistivity parameter - joint region [x$10^{-12}$ $\Omega \cdot m^2$] | 122 | 42 |
| Strand-to-joint resistivity parameter - joint region [x$10^{-12}$ $\Omega \cdot m^2$] | 2100 | 97 |
| Solder layer thickness [m] | 0 | 0 |

To account for the reduced interstrand resistances in the PFIS joints due to the higher compaction rate produced by the swaging operation, the joint interstrand resistivity parameter has been obtained matching the measured interstrand resistance distribution in cable sections in the virgin state (see [20]). Since no data is available for the terminations of the leg without wraps, the PFIS NW strand to joint resistivity parameter has been obtained by rescaling the value found for PFIS W on the base of the experimental ratio between inter-petal resistance measurements for the two conductors in the virgin state. The partial wraps removal from the outer surface of the cable in the terminations of PFISW has also been taken into account in the rescaling operation, leading to a slight reduction of the difference between the two legs. Regarding the soldering of the joints, although the use of solder is reported for the PFIS, its thickness has been neglected in the simulation due to the experimentally observed poor bonding [14].

The parameter set of PFIS W has also been used in the simulations of PFCI and PFCN1 since no specific interstrand resistance measurements exist for these conductors and similar resistivity values are expected.

The presence of Cu strands in PFCN1 affects the cable resistance reducing contacts between superconducting strands. This geometrical effect is reproduced in JackPot through the adoption of an appropriate cabling function, resulting in strand trajectories characterized by smaller contact areas than for the PFIS and PFCI. As a consequence, no significant differences are to be expected between the samples in terms of resistivity parameters, being their values connected to the choice of the materials and not to the cabling pattern.

Following the adoption of a different soldering technique in the PF conductor insert, it is likely that a non-zero solder layer is present in the joints of the sample, though no information is available regarding the possible thickness. On this account the solder layer is assumed negligible in the calculations, after having verified that it would not produce in any case significant effects on the final result. Figure 4 shows the simulated strand current distributions at the instant of quench for PFCI critical current Run # 027-02 in case of non-soldered joints. A 1.5 mm solder layer corresponding to two strand diameters, which can be regarded as a realistic maximum average thickness on the ground of the estimations of [7], was also implemented for the same run to verify the consequences of a solder layer. The peak field and the temperature of the run were 6T and 5.8 K, respectively. The addition of solder (not shown in Figure 4) has no appreciable effect on the strand current, which is mainly driven by the joint layout combined with the magnetic field and temperature profiles along the coil.

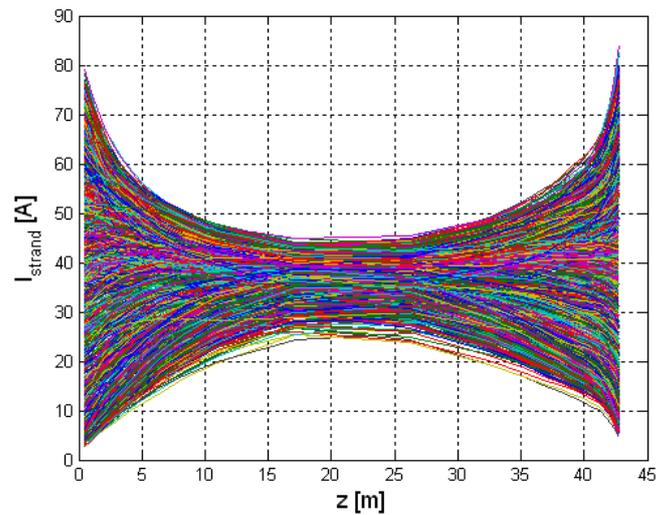

**Figure 4.** Strand current distribution along the PFCI for Run # 027-02 without soldered joints.

To make a realistic simulation of the PFCN1 SULTAN sample, the copper sole of the upper termination has been accurately simulated with FEM software coupled to JackPot, see Figure 5. Similarly to its assembly procedure, the sub-bundles from the final cabling stage (petals) in the model were "unwound" and individually connected to one of the available slots in the sole. The routine to calculate the strand connections to the joint for PFCN1 is similar to the one used in the previous joint models. A series of simulations were carried out where the petals were placed in different slots, to cover the uncertainty of how the petals were connected in the real SULTAN sample.

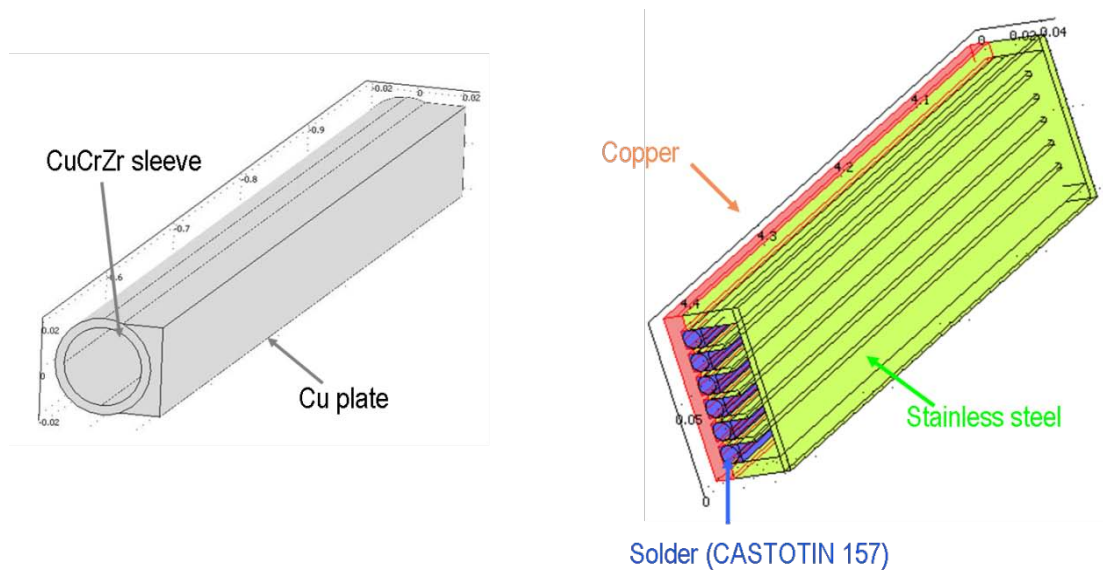

**Figure 5.** Model of the PFIS/PFCI (left) and the PFCN1 (right) terminations realised in FEM software.

JackPot usually assumes a uniform temperature everywhere in a CICC, both in a cross section and along the axial length. Although being acceptable for short samples such as in SULTAN, where the high field region has a total length of about 0.5 m, this approximation is not adequate for long coils such as the PFCI. The issue is of particular relevance for NbTi CICCs, due to their strong critical current dependence on temperature. Moreover, the PFCI test configuration featured a double cooling circuit as illustrated in Figure 6, which generally resulted in a slightly lower temperature in the conductor section close to the intermediate joint compared to the middle of the main winding. In order to take these factors into account, a longitudinal temperature profile has been implemented in the PFCI simulations through a linear interpolation of the temperature measurements at inlet, outlet and middle of the main winding (sensors Tcin, T04H and T03H in Figure 6), as illustrated in Figure 7.

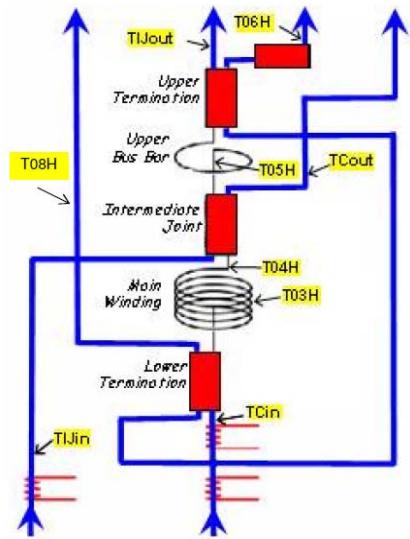

**Figure 6.** PFCI cooling scheme and temperature sensor location, courtesy of R. Zanino [6].

As an example, assuming a uniform longitudinal temperature would lead in case of Run # 035-01 (a current sharing temperature test with a transport current of 55 kA and peak background field of 6 T) to an unrealistically huge dissipation close to the intermediate joint due to the combination of current unbalance and magnetic field. However, the best available approximation of the axial temperature gradient, following the recorded data of the temperature sensors, sets the peak dissipation to its correct location close to the middle of the coil, where it was experimentally recorded, see Figure 7.

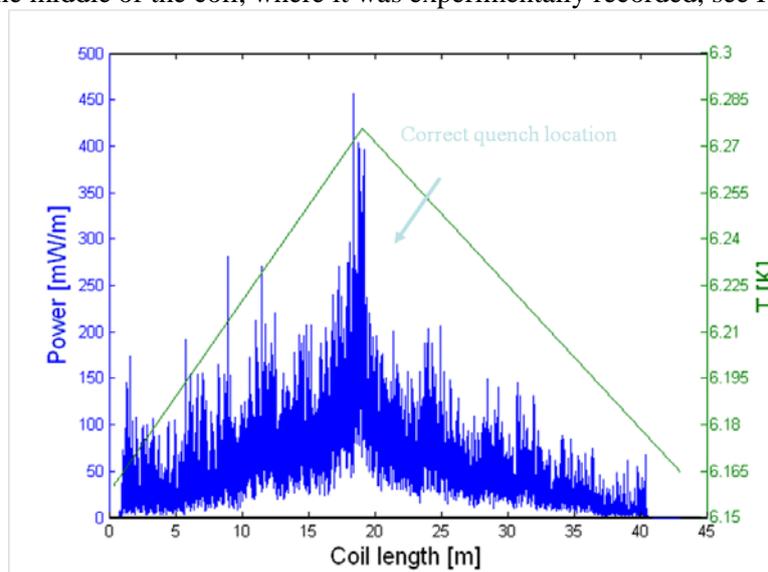

**Figure 7.** PFCI Run # 035-01 power dissipation distribution at quench taking into account the coil axial temperature profile (Y secondary axis).

4. **Validation of the simulation results**

The simulation results have been validated through the comparison with measurements for both critical current (Ic) and current sharing temperature (Tcs) runs for all the analysed samples.

Figure 8 shows a summary of the measured and calculated critical currents for different SULTAN background fields for the two legs of PFIS. The agreement between simulation and experiment is found to be good at all considered points, within 0.12 K and few kA for Tcs and Ic runs respectively. Due to the observed tendency of PFCN1 to quench even at relatively low currents without appreciable transition, critical current and current sharing temperature could be measured only for few experimental runs, which was not enough to draw an analogous graph for this conductor. However, even for PFCN1 the code is able to properly describe the behaviour of the sample, as illustrated in Figure 9 where the simulated and measured evolutions of the electric field during a Tcs run are compared. In Figure 8 and Figure 9 the similarity between the slopes of the simulated and experimental curves should be noticed, indicating that the current unbalance of the sample could be realistically reproduced by the code thanks to the detailed combined modelling of the cable and joints.

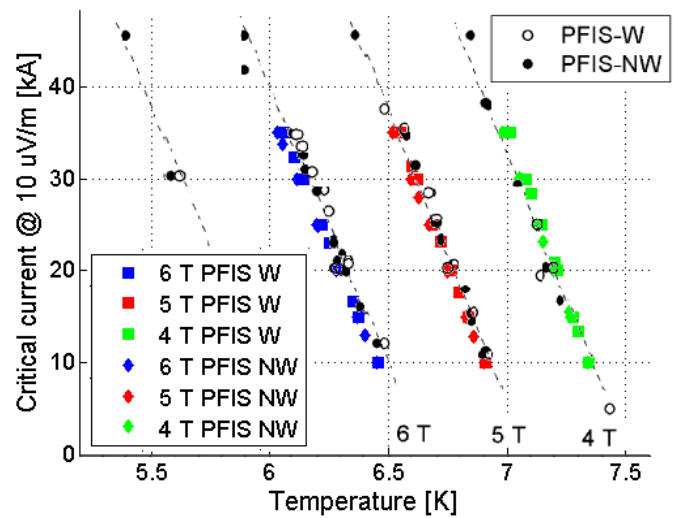

**Figure 8.** Summary of the simulated (square & diamond) and measured (dot) Ic and Tcs for PFIS.

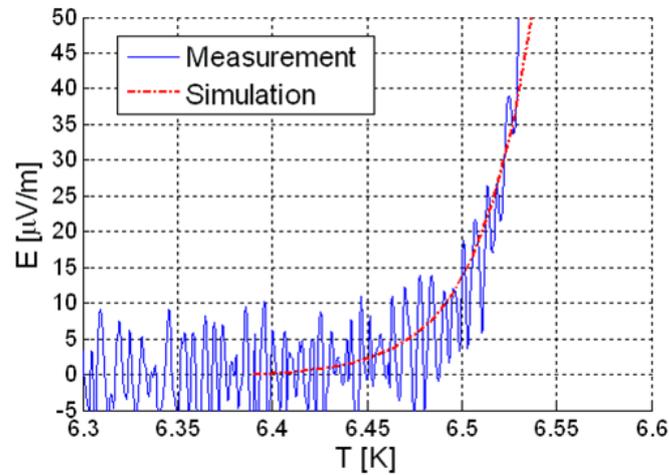

**Figure 9.** Electric field evolution versus temperature for PFCN1 Tcs run #SSPF2D180510 with transport current 20 kA and SULTAN background field 5 T, showing accurate computation of the VT trace until runaway.

The comparison between simulation and measurement appears to be somewhat less precise for the PFCI. In particular the simulation is found to generally shift the instant of quench of the cable, as shown in Figure 10. In Tcs runs, the temperature at which the coil reaches the quench electric field is on average 80 mK higher in the simulation than in the measurements. In spite of the increased difference compared to the other analysed samples, the PFCI simulation results are already much better than the predictions presented in [8, 9], where the deviation between prediction and experiment amounts up to more than 0.4 K. Since JackPot has already been proved effective in describing the behaviors of the other analyzed samples, the observed increased error cannot be simplistically attributed to limitations of the code. Instead other factors such as the length of the PFCI winding of several tens of meters combined with a rather poor axial temperature monitoring should be taken into account. The implementation of a thermo-hydraulic routine in JackPot foreseen for the near future will also be applied to obtain a final quantitative assessement of the problem current 45 kA and background field 6 T. Just below 6.2 K the measured electric field starts to deviate from the computed trace and a quench occurs at 6.25 K. In the simulation the quench electric field is shifted by 0.14 K due to thermal runaway, which is not included in the modelling.

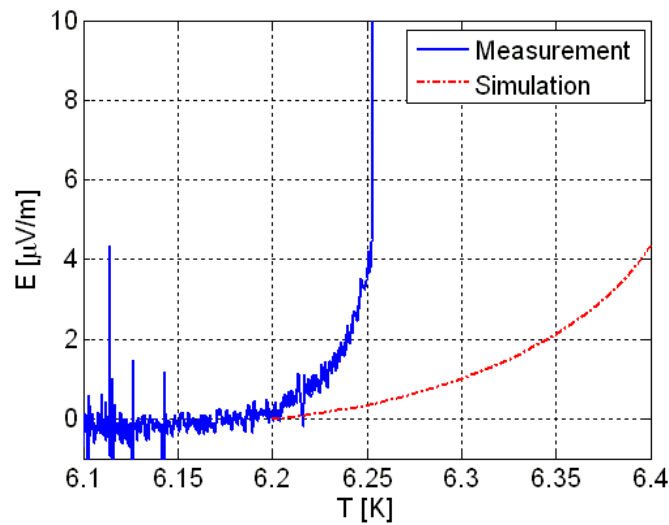

**Figure 10.** Electric field evolution versus temperature for PFCI Tcs run #157-01 with transport

5. **Dissipation spot length**

The power dissipation inside a CICC at the beginning of a quench is in the form of local dissipation spots along the strands. The length over which a dissipation spot extends depends on the properties of the strand (n-value, Ic(B) steepness) and cable (twist pitch, void fraction, diameter) and can be determined by plotting the strand peak power normalized to different strand volumes (i.e. dissipation spot lengths). If the normalization length is lower or equal than the dissipation spot extension a constant power density is found. However, when the normalization length exceeds the dissipation spot extension, the power density drastically drops as shown in Figure 11 and Figure 12. Since they are wounded out of the same cable, the dissipation spot lengths of PFIS and PFCI are found, not surprisingly, to coincide (~ 8.4 mm). The dissipation spot length of the PFCN1 is instead lower (~ 4.0 mm), which is likely to be connected to the sharper V-I transition of its strands [27] and the different conductor geometry. The dissipation spot lengths have been used in the following analysis to normalize the strand power dissipation at quench of the different samples.

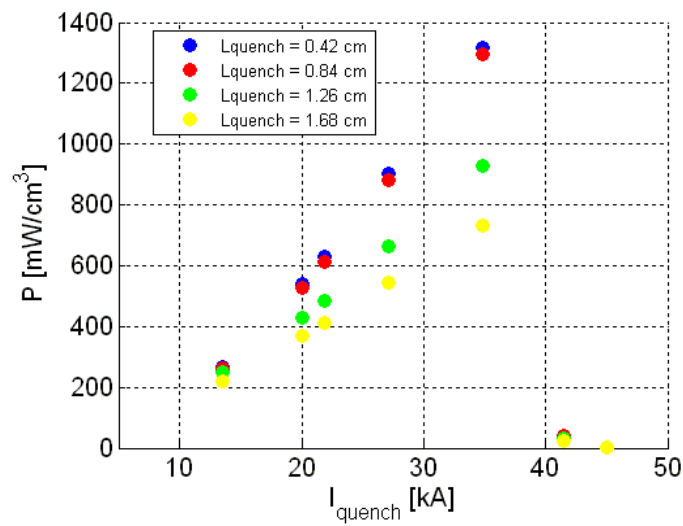

**Figure 11.** Strand dissipation versus quench current for different quench lengths of the PFIS.

13 October 2011

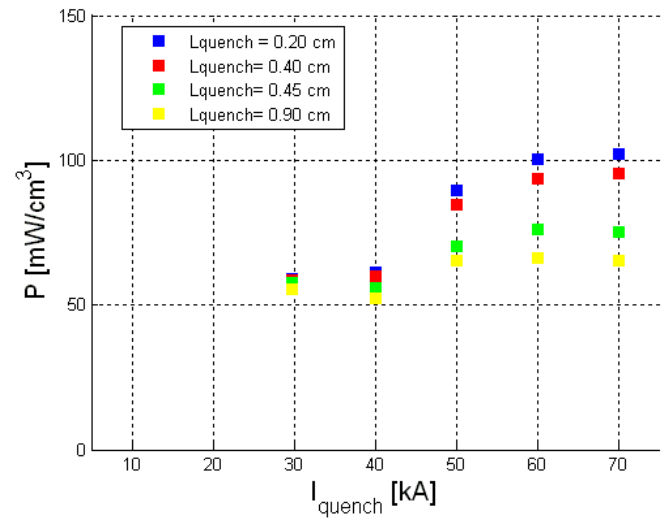

**Figure 12.** Strand dissipation versus quench current for different quench lengths of the PFCN1.

### 6. Quench power distribution

Following the precise trajectories of all the single strands inside a CICC, JackPot is the first and only available code at present to allow the visualization and quantification of the dissipation spots inside a sample although approximating it without thermal modeling. Figure 13 and Figure14 show the power dissipation at quench conditions of the strands of PFIS W at the cross section where the strand peak power density is located. It can be observed that while only few strands show significant dissipation at 41.5 kA, the power generation is spread over a wider part of the cable cross section at low currents. Moreover, in both cases the highest power dissipation takes place at x ~ rout where the peak field resulting from the sum of background and self field is located.

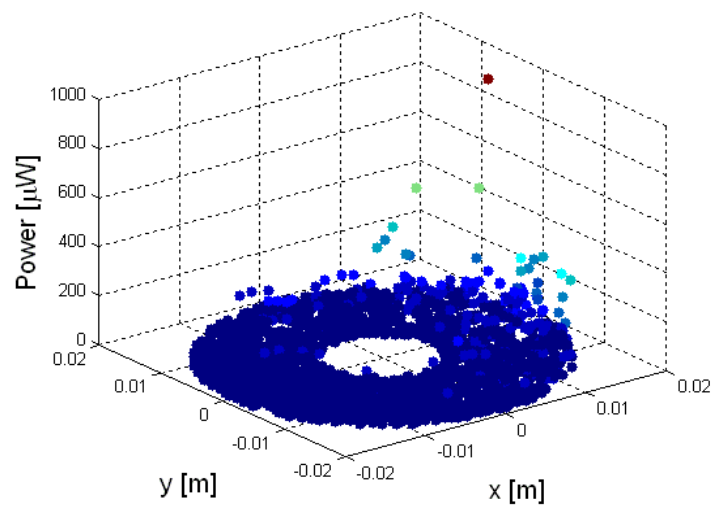

**Figure 13.** Strand dissipation distribution at quench in the cross section of PFIS W where the strand peak power generation is located for 13.5 kA at 6T.

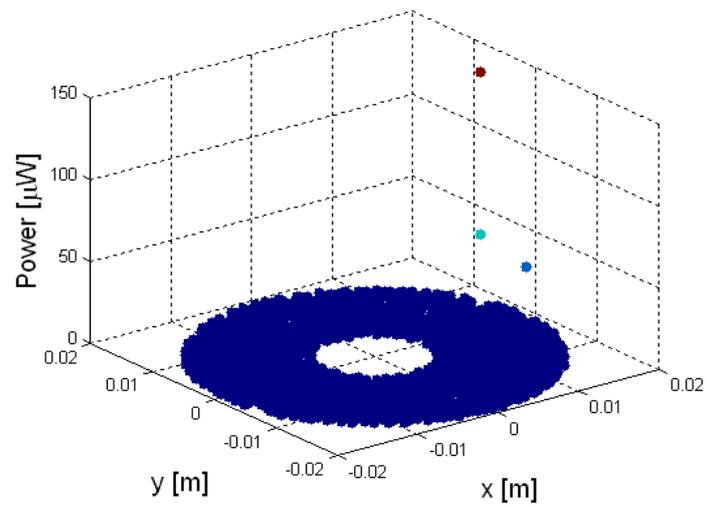

**Figure 14.** Strand dissipation distribution at quench in the cross section of PFIS W where the strand peak power generation is located for 41.5 kA at 6T.

The axial strand power distribution for the same runs of PFISW is instead illustrated in Figure15 and Figure 16. Also in this direction the number of strands taking part in the power generation at quench reduces with increasing transport current. Similar behaviours for both the strand dissipation distribution in the cross section and along the longitudinal direction are observed for the other analyzed samples.

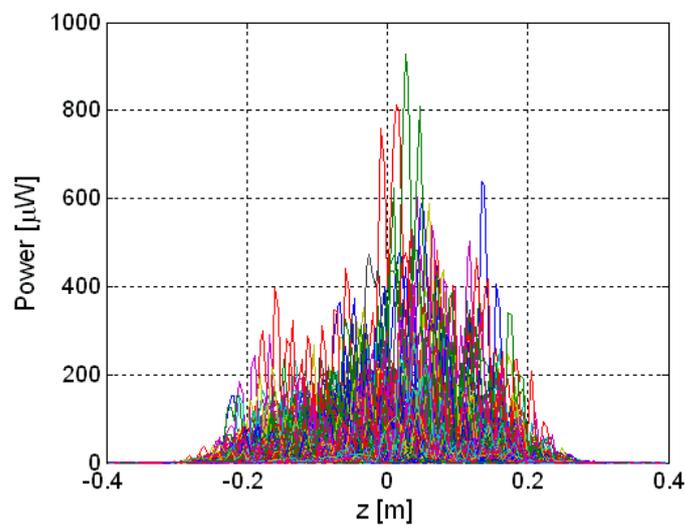

**Figure 15.** Strand dissipation distribution along the z-axis of PFIS W for 13.5 kA at 6T.

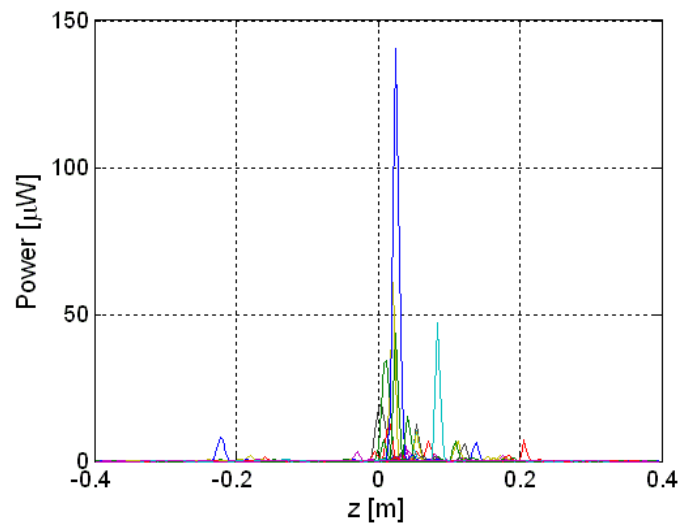

**Figure 16.** Strand dissipation distribution along the z-axis of PFIS W for 41.5 kA at 6T.

In Figure15 and Figure 16 the effect of the joint layout on the current distribution among the strands, and therefore on the power dissipation distribution can also be clearly observed. Although the high field region of SULTAN extends approximately between -0.2 m and 0.2 m in the z direction and features an almost uniform magnetic field, the resulting strand axial dissipation distribution is not symmetric to the high field region centre (i.e. z = 0 m) as it would be expected in the case of uniform current distribution among the strands. The observed lack of symmetry for PFIS W is a consequence of a number of factors, namely joint design, wraps, self field and twist pitch. The total strand power distribution along the PFIS is illustrated for both legs of the sample in the same operating conditions in Figure 17. The power distribution resulting from a hypothetical rotation of 180º of the copper plates in both terminations of PFIS W (without variation of the peak field position) is also included to show the effect of the joint layout on the power distribution. In the PFIS W the joint design, specifically the limited contact angle sleeve-Cu plate, its length and position with respect to peak field side of the conductor combined with the twist pitch sequence produce a current unbalance among the strands that can be hardly redistributed due to the presence of wraps. The configuration is such that the overloaded strands enter the high field region between z = 0-0.1 m causing an asymmetric dissipation distribution. Removing the wraps in the joint proves to be effective in reducing the current unbalance; the power dissipation of the PFIS NW sample is more symmetric than that of the PFIS W. The simple rotation of the connection plates at the extremities of the sample leg is also found to help in achieving a more uniform power spread, which demonstrates the importance of the joint design on the conductor performance, especially for short samples as the ones tested in SULTAN.

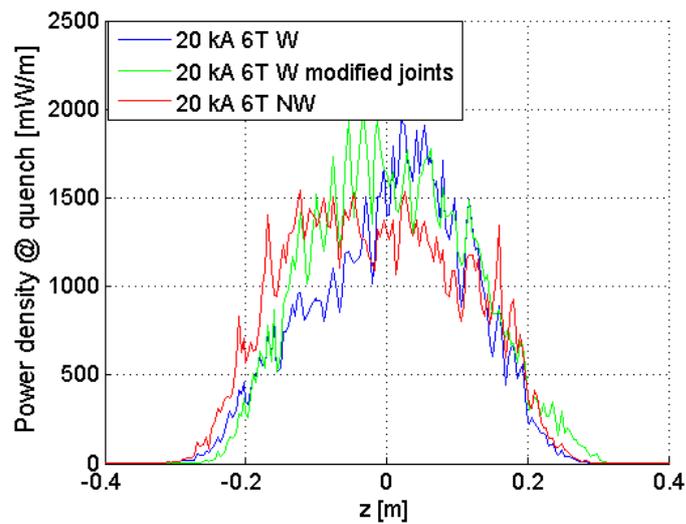

**Figure 17.** Strand power dissipation distribution along PFIS z-axis.

As extensively discussed in the introduction section, since JackPot is a pure electromagnetic model and does not include a thermo-hydraulic description of the CICC, it cannot be used to predict quenches and describe their evolution. However, when the experimental quench point is known in terms of quench current, temperature or electric field, the instantaneous power dissipation can be computed. The power densities generated in the three samples in the entire cable and in the strand with the peak dissipation at different quench currents are illustrated in Figure 18 and Figure 19. In the analysis the quench point has been defined in terms of the experimentally measured temperature or electric field at take off. An error has been also taken into account corresponding to the uncertainty in the take off instant definition.

13 October 2011

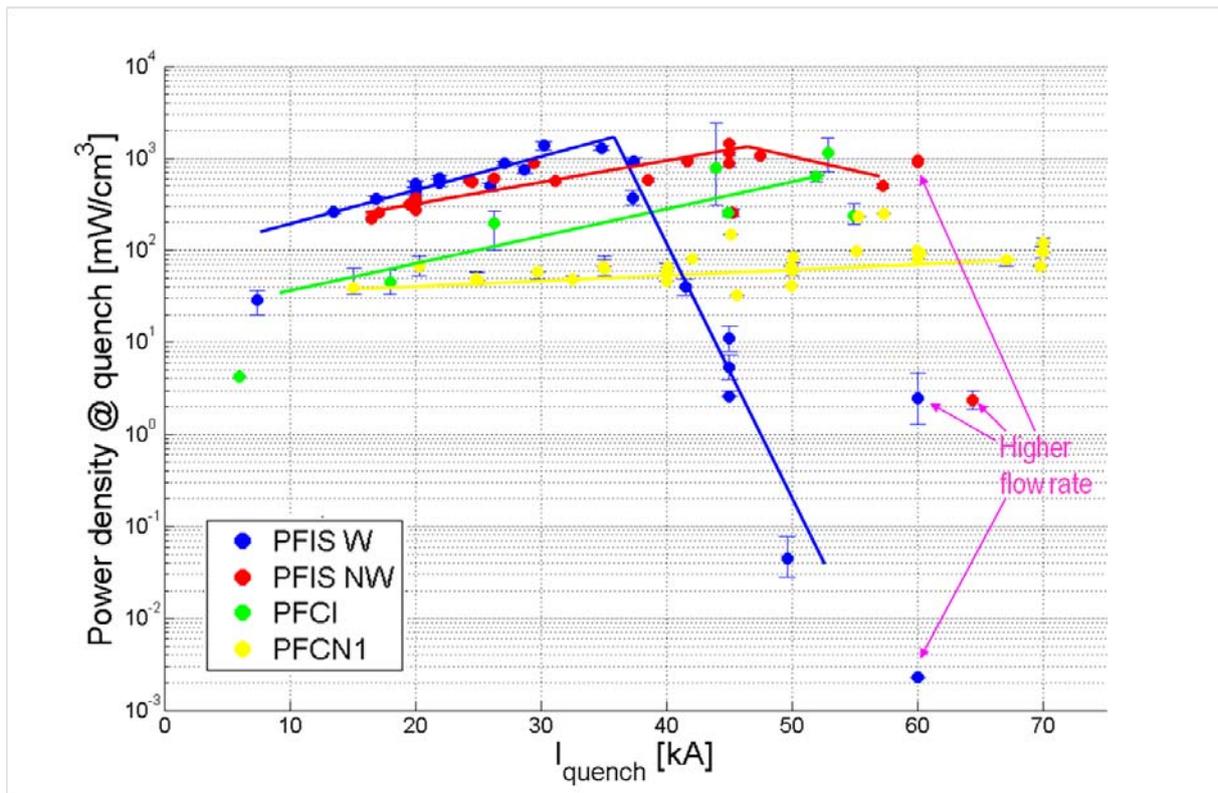

**Figure 18.** Strand peak power dissipation density at quench. Solid lines show the general behaviour of the samples.

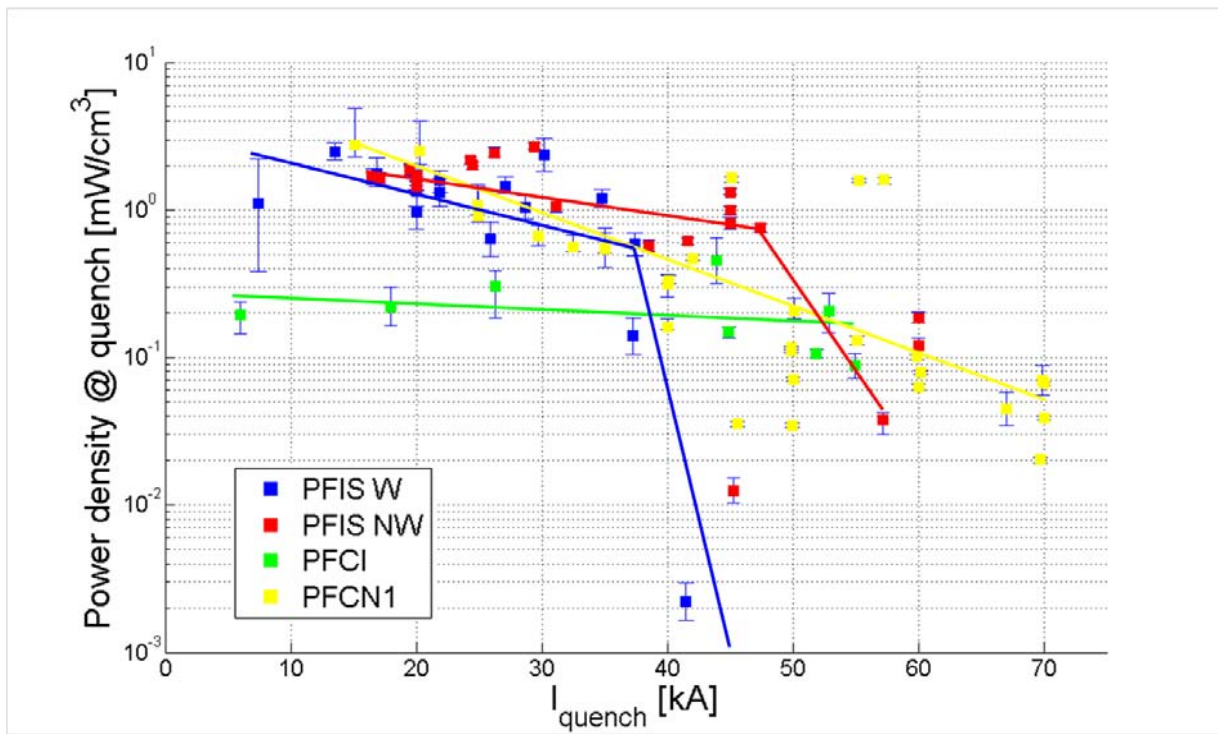

**Figure 19.** Cable power dissipation density at quench. Solid lines show the general behaviour of the samples.

The points in Figure 18 and Figure 19 have been obtained from simulations performed at different magnet field and temperature values along the entire current range explored during the experimental tests. From the plots a clear correlation among the simulations of each sample can be observed, which has been represented through solid lines.

In the PFIS both cable and strand peak power dissipation densities evolve in a similar way, dropping at 35-40 kA for the leg with wraps, while a stable behaviour is maintained in the leg without wraps up to higher currents (~ 45 kA). These current levels correspond to the experimentally observed thresholds for sudden quenches in the samples [1]. The knee in the strand power dissipation at quench of the PFIS is interpreted in terms of different quench regimes (cable and strand stability) at low and high currents as already reported for Rutherford and triplex cables as well as CICCs [24-26]. In triplex cables, increasing current non-uniformity has also been associated to a reduced current threshold for quench regime transition. The same behaviour is also seen in the PFIS where, in the left leg, the presence of wraps highly impedes current redistribution, therefore causing an earlier transition to the strand quench behaviour. The scarcity of measurements at high currents combined with the adoption of changing test conditions (ramp rate and He flow rate) for a significant number of test points above the knee make it not possible to draw further conclusions from the available data for the PFIS. As additional remark, the idea that the knee may result from an underestimation of the quenching average cable / strand temperature, increasing with cable current, seems implausible. There are no arguments to support the sudden appearance of such a strong temperature measurement error. Only a change in quench regime (transition to strand quench regime) would be likely to produce a sharper increase of the quenching strand temperature, as this would also be associated to a reduced quench power. Moreover, it would not solve the knee in the cable power dissipation density, for which the measured average strand temperature is the only available practical and valid value.The cable and strand peak power dissipation of the PFCI follow trends analogous to what observed for the PFIS in the cable quench regime. The one order of magnitude difference in the strand peak power density of this sample indicates better current redistribution capability. The lower cable dissipation is, instead, to be attributed to the different B-field gradient of the PFCI giving a not perfectly uniform magnet field in the high field region of the sample in combination with the temperature gradient existing along the winding. No clear sign of quench regime transition could be found for the PFCI, possibly because it lies beyond the tested experimental range.

The cable power dissipation of the PFCN1 is higher than that of the PFCI, but in the same range of the PFIS one since the two samples feature similar magnet field and temperature gradients in their high field regions. However, the strand peak power dissipation at quench is much lower due to a better current distribution among the strands and, unlike the PFIS samples and PFCI, remains quite constant in the whole tested range.

An interesting observation is that the PFIS cables could actually reach higher local peak power densities than the PFCN1 and to a less extent than the PFCI. A correlation with the level of current non-uniformity seems credible and is discussed in the next section.

7. **Current unbalance**
The claim that current unbalance is the main driver for the early quenches in the PFIS cable is justified with Figure 20, showing the ratio between the current flowing in the strand with the peak power dissipation and the average strand current at the instant of quench, as derived from the JackPot results.

Although they remain stable around 2.2 and 2.0 for PFIS W and PFIS NW, respectively, up to 30 kA, the strand overloads increase for higher quench currents which leads in the end to the appearance of premature quenches. The increase of the unbalance in this sample was already indirectly foreseen in [5] on the ground of the growing difference between experimental quench current and peak field critical current estimation. Since the threshold for the overload rise does not coincide with the experimental quench regime transition current levels previously identified, no clear quantitative correlation could be determined between strand overload and premature quench of the conductor at

13 October 2011

present. Therefore the analysis suggests that no comprehensive conclusions can be drawn focusing exclusively on the current unbalance, but other factors may need to be taken into account. In this respect, the planned addition of a thermal-hydraulic routine may help to better quantify the current overload and peak dissipation relationship. Regarding the PFCI, Figure 20 shows that the strand overload at quench is much lower (~ 1.3-1.4) and hardly varies with quench current. Since the coil featured the same strand, cable and joint design as the PFIS, the improvement of the current unbalance is merely related to the higher distance between joints and high field region of the winding, which allows a better re-distribution of the non-uniformity.

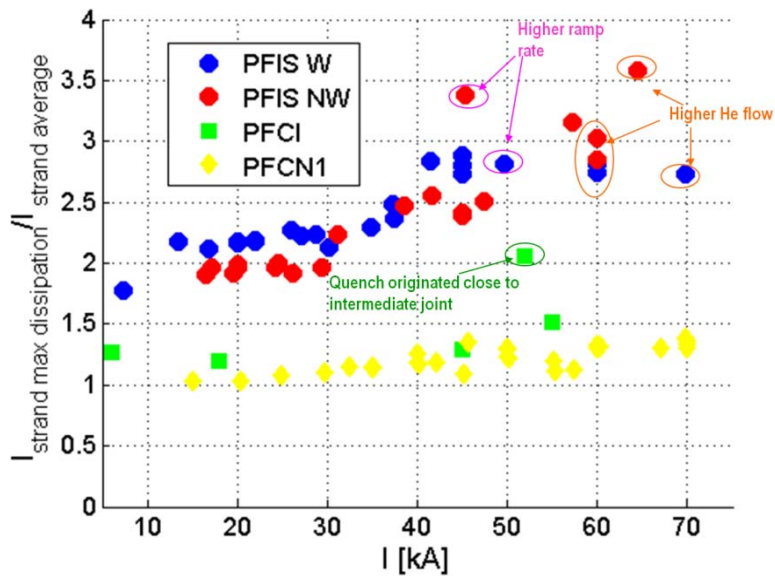

**Figure 20.** Overload of the strand with peak power dissipation at quench versus quench current.

For only one point in the PFCI data, at 52 kA, the strand overload is found to be comparable to the values of PFIS. However, the result of this particular run has been incorrectly localised in the cable. Although the measurements show that the quench originates in the middle of the main winding, JackPot predicts the quench to start close to the intermediate joint due to the combination of elevated non-uniformity produced by the joint design and quite strong magnet field. A more detailed knowledge of the temperature profile along the coil would be essential for improving the definition of the quench origin position in the simulation of this data point but it is not available.

Although the PFIS test campaign also included a number of tests with a higher current ramp rate and helium flow rate, their number is too small for drawing any conclusions.

The final sample (PFCN1) shows the lowest and most stable current unbalance of all the analyzed conductors. Since the PFCN1 is a short sample like the PFIS, no significant current re-distribution is expected between joint and high field region due to the short intermediate distance. The improved current uniformity therefore comes from the new layout employed for the upper terminations and from the exclusion of the bottom joint.

Table 6 provides a summary of the peak power at quench and current non-uniformity level at operating conditions for the analyzed samples.

**Table 6.** Peak quench power and current non-uniformity of the samples at operating conditions of ITER PF coils.

|  | PFIS W | PFIS NW | PFCI | PFCN1 |
|---|---|---|---|---|
| Type | PF1 | PF1 | PF1 | PF2 |
| Nominal Operating Current [kA] | 52.0 | 52.0 | 52.0 | 50.0 |
| Nominal Effective Peak Field [T] | 5.7 | 5.7 | 5.7 | 5.0 |
| Strand peak power at quench [mW/cm$^3$] | 0.02 | 800 | 640 | 65 |
| Cable power at quench [mW/cm$^3$] | 0.8e-6 | 0.4 | 0.15 | 0.21 |
| Current non-uniformity at quench | 3.05 | 2.8 | 1.45 | 1.25 |

As already observed, the simulation results show that the PFIS cables could reach higher local peak powers than the PFCN1 and, to a less extent, than the PFCI. Although speculative and not supported by the analysis presented here, the most likely explanation for this behaviour is that the local higher peak power in the PFIS compared to the PFCN1 could possibly be explained by a larger temperature gradient and thus better cooling of the hot spots by their immediate surrounding. In the case of large current non-uniformity, the heat in the cable volume is concentrated in very small spots with much higher temperature than most of the surrounding strand volume characterized by a much lower dissipation. This surrounding cable volume at lower temperature offers better cooling conditions than in cables with more homogeneously distributed power dissipation. This possible thermal phenomenon cannot be clearly separated from the so-called cable and strand quench regimes as observed for the PFIS, with high quench power at low current and above a certain current threshold, decreasing quench power with higher current. This regime separation is often explained as the possibility for single strands to recover from initial quenching through interaction with other strands (cable) and the impossibility for such interaction above the current threshold. In addition it should be remarked that the PFCN1 cable is made of different strands, and the cable has a different configuration.

To summarize, the layout adopted for the PFIS has been proved to cause high current non-uniformity among the strands. The presence of wraps in the left leg of PFIS together with the short distance between the terminations of the sample and the high field region make current re-distribution hardly possible. As a consequence of the increasing current unbalance, this sample enters the strand quench regime beyond 38 kA, where the available cooling becomes insufficient to remove the heat generated by the overloaded strands, and the cable quenches before reaching the nominal operating current of 52 kA. In the right leg of the PFIS, the absence of wraps slightly improves the strand current uniformity, moving the threshold for the transition to the strand quench regime to 45 kA. Since the joints of the PFIS and the PFCI are quite similar, the better performance of the latter can be attributed to the much larger distance between its joints and the high field region. Here, the current unbalance originated at the joints can be reduced to a low enough level to prevent a quench from happening due to overloaded strands. In this case an important role is also played by the cooling circuit layout of the coil, which allows for a reduced temperature close to the Intermediate Joint (see Figure 7), preventing a quench at this location where the unbalance level is still high. Finally an effective way for attaining low current non-uniformity even in short samples is demonstrated with the PFCN1, where a nearly perfectly homogenous and stable current distribution is obtained through the elimination of the bottom joint and the individual solder of the petals in the upper terminations. The resulting strand peak power generation at quench is found to be not only one order of magnitude lower than in the other samples, but also practically constant in the whole tested range. The simulations show the PFCN1 to be characterized by a small current overload, even somewhat lower than what found for the PFCI, and as for this sample, no sign of quench regime transition is observed.

The influence of the sample and joint layout on the current distribution and its relation with strand and cable peak power at quench have been quantitatively demonstrated for the short sample and insert coil testing of ITER PF conductors. The importance of this analysis lays in quantitatively specifying the boundaries for the design of the ITER PF coils within the anticipated coil-joint configuration, joint properties and applied magnet field conditions. The strand power dissipation at quench sets a clear limit to the allowed current non-uniformity and thus to the joint properties in the sense of resistivity of the used components, which are chosen to limit heat dissipation originated by AC losses from applied AC magnet field variations.

8. **Conclusions**

For the short sample and insert coil testing of ITER PF conductors we have quantitatively determined the influence of the sample and joint layout on the current distribution and its relation with strand and cable peak power at quench with the JackPot code. Although lacking of a thermal-hydraulic description of the CICC, the model allows the computation of the instantaneous power dissipated in the strands, once the quench conditions (current, temperature and magnetic field) are known from the experiment. JackPot allowed the determination of the $T_{cs}$ of the analysed samples in all experimental conditions with a maximum error of 0.12 K, which is up to now is a great improvement in accuracy compared to other existing models. Moreover, for the first time the detailed power dissipation distribution and current non-uniformity of the analysed samples could be quantified and visualized.

According to the model, a quench in large NbTi CICCs is initiated in relatively small isolated hot spots often composed by single wires with peak dissipation along lengths of less than 1 cm. The hot spots are located in the high magnetic self field zone of the cable cross section. In case of elevated current non-uniformity among the strands, the peak power dissipation increases with cable current.

For the PFIS, the cable and strand peak power dissipations evolve similarly, showing drops at the same current levels at which sudden quenches start being observed in the experiments. The behaviour is explained in terms of two different quench regimes: at low currents, the quench is determined by the global cable behaviour, whereas at high currents, localized power dissipation on single strand level seems responsible for the quench. The strand quench regime transition is not observed for PFCI and PFCN1; both samples show reduced strand peak power densities at quench due to more homogeneous current distribution. The improved current distributions in the latter two samples are caused by either longer distance between the joints and the high field region (PFCI), or an improved joint design (PFCN1). No clear quantitative correlation could be verified between the current unbalance of PFIS and the quench regime transition, although the constantly increasing strand current overload can be regarded as the origin of the observed premature quenches. However, based on the experimentally measured temperature profile, the voltage-temperature characteristics could be calculated with high accuracy along the I, B and T spectrum of the experiments thanks to the detailed level of joint and strand current simulation. For the actual runaway process, other variables such as the strand temperature distribution obviously play a role. In this sense, the addition of a thermal-hydraulic routine is the natural following step in the analysis in order to gain a better insight into the process.

From the simulation results, a maximum current unbalance of 1.45 at operating conditions is allowed in ITER PF cables to avoid the transition to the quench strand regime, and therefore prevent the risk of sudden quenches. In these conditions the expected cable power dissipation at quench is in the range of 0.15-0.25 mW/cm$^3$; while the strand peak power varies, depending on the strand characteristics and cable configuration, between 65-640 mW/cm$^3$.

The initial assumption that by first order approximation the small scale and local electrical-thermal strand interactions were not unambiguously required for a pragmatic performance analysis appeared justified as the influence of these effects in the modeling were identified as being marginal and far less than the generally accepted temperature margin for safe operation of large sized NbTi CICCs. This means that the approach is allowed to be used for joint design and setting the safety limits for peak power dissipation and maximum current non-uniformity.